\documentclass[useAMS,usenatbib,usegraphicx,letterpaper]{mn2e}
\usepackage{color}
\usepackage{times}
\usepackage{hyperref}
\usepackage{xspace}
\usepackage{graphicx}
\usepackage[utf8]{inputenc}
\usepackage[T1]{fontenc}
\usepackage{aecompl}
\usepackage{pslatex}
\usepackage{textcomp}
\usepackage{footnote}
\usepackage{enumitem}
\usepackage[portuguese,english]{babel}

\hypersetup{
    pdftitle={Magnified SN}, %
    pdfauthor={Jakob Nordin},
    colorlinks=true,        %
    linkcolor=blue,         %
    citecolor=blue,         %
    urlcolor=blue           %
}

\newcommand{\NOTE}[1]{}                       %

\newcommand{\new}[1]{{#1}} %
\newcommand{\todo}[1]{}
\newcommand{\ny}[1]{{#1}} %
\newcommand{\neu}[1]{{#1}}
\newcommand{\refreply}[1]{{#1}}

\newcommand{\nodata}{...}

\title[Lensed SNe as Probes of Cluster Mass Models]{\new{Lensed Type Ia Supernovae as Probes of Cluster Mass Models
}
}

\author[Nordin et al.]{
{\parbox{\textwidth}{J. Nordin{$^{1, 2}$},
D. Rubin{$^{3}$},
J. Richard{$^{4}$},
E. Rykoff{$^{5}$},
G. Aldering{$^{1}$},
R. Amanullah{$^{6, 7}$},
H. Atek{$^{8}$},
K. Barbary{$^{9}$},
S. Deustua{$^{10}$},
H. K. Fakhouri{$^{1, 11}$},
A. S. Fruchter{$^{10}$},
A. Goobar{$^{6, 7}$},
I. Hook{$^{12, 13}$},
E. Y. Hsiao{$^{14}$},
X. Huang{$^{11, 15}$},
J.-P. Kneib{$^{8, 16}$},
C. Lidman{$^{17}$},
J. Meyers{$^{18}$},
S. Perlmutter{$^{1, 11}$},
C. Saunders{$^{1, 11}$},
A. L. Spadafora{$^{1}$},
N. Suzuki{$^{19}$}
}
}
\\
\\{\Large (The Supernova Cosmology Project)}\\
\\
{\parbox{\textwidth}{$^{1}${E.O. Lawrence Berkeley National Lab, 1 Cyclotron Rd., Berkeley, CA, 94720; jnordin@lbl.gov\\}
$^{2}${Space Sciences Lab, University of California Berkeley, 7 Gauss Way, Berkeley, CA 94720\\}
$^{3}${Department of Physics, Florida State University, Tallahassee, FL 32306, USA\\}
$^{4}${Centre de Recherche Astronomique de Lyon, Université Lyon 1, 9 Avenue Charles Andre, F-69230 Saint Genis Laval, France\\}
$^{5}${Kavli Institute for Particle Astrophysics and Cosmology, SLAC National Accelerator Laboratory, Menlo Park, CA 94025\\}
$^{6}${The Oskar Klein Centre, Department of Physics, AlbaNova, Stockholm University, SE-106 91 Stockholm, Sweden\\}
$^{7}${Department of Physics, Stockholm University, Albanova University Center, SE-106 91, Stockholm, Sweden\\}
$^{8}${Laboratoire d'astrophysique, École Polytechnique Fédérale de Lausanne (EPFL), Observatoire de Sauverny, 1290, Versoix, Switzerland\\}
$^{9}${Argonne National Laboratory, 9700 South Cass Avenue, Lemont, IL 60439, USA\\}
$^{10}${Space Telescope Science Institute, 3700 San Martin Drive, Baltimore, MD 21218\\}
$^{11}${Department of Physics, University of California Berkeley, Berkeley, CA 94720\\}
$^{12}${Department of Physics (Astrophysics), University of Oxford, DWB, Keble Road, Oxford OX1 3RH, UK\\}
$^{13}${INAF - Osservatorio Astronomico di Roma, via Frascati, 33, I-00040 Monte Porzio Catone, Roma, Italy\\}
$^{14}${Carnegie Observatories, Las Campanas Observatory, Colina El Pino, Casilla 601, La Serena, Chile\\}
$^{15}${University of San Francisco, 2130 Fulton Street, San Francisco, CA 94117-1080\\}
$^{16}${Aix-Marseille Université, CNRS, LAM (Laboratoire d'Astrophysique de Marseille) UMR 7326, 13388, Marseille, France\\}
$^{17}${Australian Astronomical Observatory, PO Box 296, Epping, NSW 1710, Australia\\}
$^{18}${Department of Physics, Stanford University, 450 Serra Mall Stanford, CA 94305, USA\\}
$^{19}${Kavli Institute for the Physics and Mathematics of the Universe (IPMU), University of Tokyo, 5-1-5 Kashiwanoha, Kashiwa, Chiba 277-8583, Japan}
}
}}

\begin{document}

\date{Accepted 2014 February 25. Received 2014 February 24; in original form 2013 December 8.}

\pagerange{\pageref{firstpage}--\pageref{lastpage}} \pubyear{2002}

\maketitle

\label{firstpage}

\newcommand{\psfphotlowisolated}{$\sim 8\%$\xspace}
\newcommand{\psfphotmatch}{a few mmag\xspace}
\newcommand{\wfcIRzps}{?\xspace}
\newcommand{\supernovacounts}{?\xspace}
\newcommand{\LCDM}{$\Lambda$CDM\xspace}
\newcommand{\HOneMag}{0.3\xspace}
\newcommand{\HOneSig}{$\sim 2 \sigma$\xspace}

\newcommand{\snia}{SNe~Ia\xspace}
\newcommand{\union}{\texttt{Union2.1}\xspace}

\newcommand{\snAmag}{\ensuremath{-0.17}\xspace}
\newcommand{\snAmagSALTtwotwo}{\ensuremath{-0.38}\xspace}
\newcommand{\snAmagerr}{0.18\xspace}
\newcommand{\mapAmag}{\ensuremath{-0.37\pm 0.06}\xspace}

\newcommand{\snHmag}{\ensuremath{-0.11}\xspace}
\newcommand{\snHmagerr}{0.14\xspace}
\newcommand{\mapHmag}{\ensuremath{-0.36 \pm 0.05}\xspace}
\newcommand{\snHfainter}{0.25\xspace}
\newcommand{\snHfaintersigma}{1.6\xspace}

\newcommand{\snLmag}{\ensuremath{-0.73}\xspace}
\newcommand{\snLmagerr}{0.14\xspace}
\newcommand{\mapLmag}{\ensuremath{-0.38\pm 0.08}\xspace}
\newcommand{\snLbrigher}{0.35\xspace}
\newcommand{\snLbrighersigma}{2.1\xspace}

\newcommand{\brightzps}{25.630, 26.082, 25.352, 25.401, and 24.710\xspace}
\newcommand{\faintzps}{25.600, 26.052, 25.322, 25.371, and 24.680\xspace}

\begin{abstract}

Using three magnified Type Ia supernovae (SNe Ia) detected behind CLASH clusters, we perform a first pilot study to see whether standardizable candles can be used to calibrate cluster mass maps created from strong lensing observations. Such calibrations will be crucial when next generation HST cluster surveys (e.g. FRONTIER) provide magnification maps that will, in turn, form the basis for the exploration of the high redshift Universe. \refreply{We classify SNe using combined photometric and spectroscopic observations, finding two of the three to be clearly of type SN Ia and the third probable.} The SNe exhibit significant amplification, up to a factor of 1.7 at $\sim5\sigma$ significance (SN-L2). We conducted this as a blind study to avoid fine tuning of parameters, finding a mean amplification difference between SNe and the cluster lensing models of $0.09 \pm 0.09^{stat} \pm 0.05^{sys}$ mag. \refreply{This impressive agreement} suggests no tension between cluster mass models and high redshift standardized SNe Ia. However, the measured statistical dispersion of $\sigma_{\mu}=0.21$ mag appeared large compared to the dispersion expected based on statistical uncertainties ($0.14$). Further work with the supernova and cluster lensing models, post unblinding, reduced the measured dispersion to $\sigma_{\mu}=0.12$. An explicit choice should thus be made as to whether SNe are used unblinded to improve the model, or blinded to test the model. As the lensed SN samples grow larger, this technique will allow improved constraints on assumptions regarding e.g. the structure of the dark matter halo.

\end{abstract}

\begin{keywords}
  Supernovae: general --- cosmology: observations --- gravitational lensing: strong --- galaxies: clusters: general --- dark matter
\end{keywords}

\newpage

\section{Introduction}\label{sec:introduction}

Clusters of galaxies are the most massive bound objects in the universe. They are dominated by their dark matter halos, which gravitationally distort and magnify background objects via gravitational lensing. This allows them to act as powerful gravitational telescopes, thereby offering unique opportunities to observe extremely distant galaxies \citep[e.g. ][]{2004ApJ...607..697K}.
Lensing magnification of up to a factor $\sim70$ (i.e. up to $\sim4.5$ mag) has been observed for multiply lensed images, and typical magnification factors of 5-10 are very common within the central one arc-minute radius of massive cluster lenses. Since the lensing amplification corresponds to a gain factor $\mu^2$ in exposure time, observations otherwise too distant and faint are made possible, opening a window to the unexplored high-redshift universe.

Today, mass maps have been constructed for many clusters, mainly relying on the positions of multiple counterparts of strongly lensed galaxies \citep[see e.g.][Richard et al. in prep.]{2010MNRAS.402L..44R,2011A&ARv..19...47K}.
Potential systematic uncertainties result from the sparse data, forcing assumptions to be made regarding physical properties. 
\neu{A well-known issue is the mass-sheet degeneracy, in which
the distortions and flux ratios from gravitational lensing
are unaffected by a change in the mean mass surface density
\citep{1985ApJ...289L...1F,1988ApJ...327..693G}.  Strongly
lensed galaxies at multiple redshifts can break this degeneracy
\citep{2004A&A...424...13B}. However, substructure within clusters can act
like localized mass sheets \citep{2012MNRAS.425.1772L,2013A&A...559A..37S},
and thus add some uncertainty to the cluster mass models. The absolute amplification, such as that measured from a standard candle, is not subject to this degeneracy and thus can be used to break it or constrain its amount \citep{1998MNRAS.296..763K}.
In addition to these physical complications, different teams may make different implementation choices, for instance in their selection criteria for multiple images.}
However, until now there has not been an independent way of testing strong lensing mass maps and their quoted uncertainties. This will be necessary in order to properly interpret findings in high magnification regions.

Each cluster observation also presents the opportunity to observe transient objects, thus potentially pushing the redshift limits for e.g. supernovae \citep{2000MNRAS.319..549S, 2003AA...405..859G}.
Ground-based searches for lensed supernovae using near-IR observations have reported two SNe behind Abell 1689: a Type IIp SN with predicted amplification $\Delta m = 1.4$ \citep{2009A&A...507...61S,2009A&A...507...71G} and a Type IIn SN with $\Delta m = 1.6$, the most amplified supernova to date provided the cluster mass model is correct \citep{2011ApJ...742L...7A}. 
However, Type II SNe exhibit a large scatter in brightness and thus cannot be used to independently measure amplification. See e.g. \citet{2002ApJ...566L..63H} for a discussion of Type II SN standardization.

We here describe a pilot study of three Type Ia supernovae ({\snia}) discovered behind clusters observed as part of the Cluster Lensing and Supernovae with Hubble \citep[CLASH;][]{2012ApJS..199...25P} programme, and how these
can be used as ``test beams'' to compare with amplifications predicted by strong lensing-based models. 
\snia~ have been used as standardized candles to detect the accelerated expansion of the Universe \citep{riess98a,perlmutter99a}, and can, with modern calibration based on the observed lightcurve shape and color, yield distance estimates with a measured scatter at the $\sim0.14$ magnitude level \citep{conley11a,union2.1}.
Although the uncertainty in lens modeling of the foreground cluster \new{adds an additional systematic error when SNe found behind clusters are used as cosmological probes}, the problem can be inverted and any changes to SN luminosity can be used to test cluster mass models or break the mass-sheet degeneracy \citep{1998MNRAS.296..763K}.
\ny{Previously such studies have only been performed using weak lensing. 
For instance, in \citet{2010MNRAS.402..526J}, the Hubble residuals of 24 \snia in the GOODS fields were compared with galaxies along the line-of-sight, providing constraints on the scaling law between velocity dispersion and galaxy luminosity.}

\new{Dark matter substructure in the cluster halo is expected to yield magnification differences around $\sim0.05$~mag \citep[see discussion on errors in well-constrained strong lensing mass models in][]{2007ApJ...668..643L,2009MNRAS.395.1319J}.
If the luminosity of SNe show discrepancies with the cluster mass model predictions, this could challenge the current assumption of no substructure.
However, the SN Ia measured dispersion is still $\sim3\times$ larger than substructure predictions, meaning that $\sim80$ SNe would be needed to confirm that estimate. \neu{Larger discrepancies, for instance due to the mass-sheet degeneracy in systems with only one strong lens, may be detectable with a much smaller sample.}
In that spirit we have undertaken this study to, for the first time, test cluster mass models using amplification.
}

In Sec.~\ref{sec:clash} we describe the CLASH survey and the modifications made in order to facilitate detection of SNe in and behind the clusters. The discovery of the lensed SNe are described in Sec.~\ref{sec:discovery}, and their lightcurves and Hubble residuals are presented in Sec.~\ref{sec:data}. The cluster mass models are presented in Sec.~\ref{sec:maps}, and the two magnification estimates are discussed in Sec.~\ref{sec:disc}. We conclude in Sec.~\ref{sec:conc}.

\ny{This study was performed blind to prevent a sub-conscious bias towards choices that agree better with the expected result.}
The analysis of the SN amplifications was kept separate from the determination of lensing maps until both were considered complete. Only after this were the derived magnitudes compared.
Additional work was done after unblinding, as described further in Sec.~\ref{sec:disc}.

\section{Clash}\label{sec:clash}

The Cluster Lensing And Supernova survey with Hubble (CLASH) program was a 524-orbit survey of 25 galaxy clusters, and was part of the Hubble Space Telescope (HST) multi-cycle-treasury programs \citep{2012ApJS..199...25P}.
Each cluster was observed with up to $16$ Advanced Camera for Surveys (ACS) optical and Wide Field Camera 3 (WFC3) IR filters for a total observation time $\sim 20$ orbits, which allowed precise photometric redshift estimates of all arcs. This is a core requirement for determination of the cluster mass profile -- a main goal of the CLASH program. Visits were separated by roughly two weeks and each cluster was monitored for $\sim3$ months. 
Simultaneously, HST observations of the parallel fields were used for a search for field SNe by the CLASH team~\citep[see e.g.][]{2012ApJ...746....5R}, where lensing effects are small and {\snia} can be used for probing dark energy and SN rate studies \citep{2007ApJ...659...98R,2008ApJ...681..462D, 2012ApJ...745...31B}. 
\new{
\citet{2013arXiv1310.3495G} recently presented eleven SN~Ia (four at $z>1.2$) detected in the CLASH parallel observations, finding rates consistent with previous high redshift studies.
}

Searching for SNe in the clusters was not part of the original CLASH survey and we proposed to find and follow these targets.
As a search using so many different filters observed in an arbitrary order will be less sensitive than one using a few dedicated search bands, we worked in coordination with the CLASH team to ensure that observations were scheduled such that the maximum SN search sensitivity was achieved while not changing the total exposure times and sequence of cameras.
First, we optimized the observing sequence to ensure that we could detect SNe.
This was performed by requiring that each observation epoch after the first
epoch must contain at least one filter that was previously observed on the
cluster, allowing us to find transients via image subtraction.
Second, because of the short time baseline on the coverage of each cluster,
edge effects were very important.  In particular, supernovae near maximum light
in the first epoch would not be discovered with sufficient time to \new{schedule any additional required observations}.  We thus placed as wide a range of wavelengths as possible in the first epoch, maximizing the chance that a supernova found after maximum light would have well-constrained colors, even without triggered follow-up.  For the following epochs, we also attempted to get as wide a range of wavelengths as possible, when compatible with the other constraints.

Given this optimized filter cadence, it was realized that background SNe amplified by gravitational lensing due to the foreground cluster could also be studied, and both teams undertook this work as well.
In order to provide full lightcurves of any SNe detected in or behind the clusters, we were granted 12 orbits of ACS and/or WFC3 observations to follow up these SNe (HST-GO: 12360).

The SNe observations described here are thus unusual in that they are based on a more diverse selection of filters than typical of the fixed bands used in all previous SN searches. 
Two of the three candidates, nevertheless, have lightcurves conforming to current SN cosmology requirements (as discussed in Sec~\ref{sec:data}).

\section{Discovery and Confirmation}\label{sec:discovery}

Built on a previous ACS cluster SN survey \citep[for details, see][]{2009AJ....138.1271D}, 
a pipeline was constructed where CLASH WFC3-IR and ACS observations were automatically downloaded, bias de-striped, charge transfer efficiency-corrected, cleaned for cosmic rays, astrometrically registered, drizzled \citep{2002PASP..114..144F}, and sky-subtracted (only the last three steps are necessary for IR images). 
Whenever an earlier observation in the same filter existed this was subtracted from the new data, and the difference image searched for suitable candidates. The last step involved a manual scan of remaining candidates (typically $\sim30$). We will here focus on our discoveries of background {\snia} lensed by the clusters.

\subsection{SN-A1 -- Abell~383}

\begin{figure}
\includegraphics[width = 3.in]{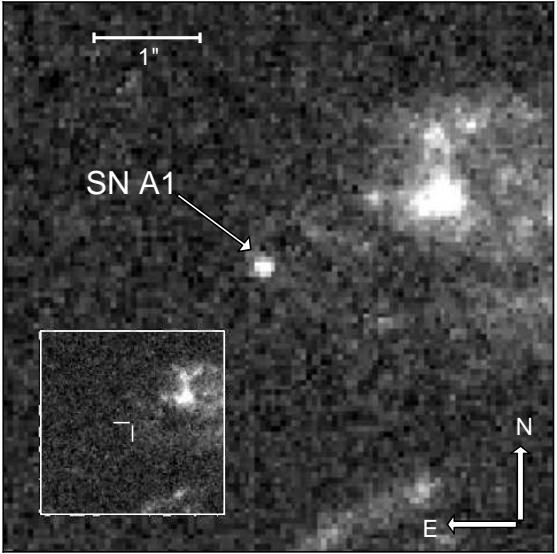}
\caption{SN-A1 behind Abell 383; the inset shows the field prior to explosion (both ACS-F814W).}
\label{fig:abell383}
\end{figure}

SN A1 was detected in the field of Abell~383 \refreply{\citep[$z=0.187$;][]{1958ApJS....3..211A}}
at $RA=42.00532$ $Dec= -3.55469$ in an ACS-F814W observation taken on Dec 28 2010 (UT). ACS-F435W did not show SN flux, making the candidate a likely high-redshift supernova. This was confirmed in subsequent ACS-F625W and ACS-F850LP observations, which all showed a good match to a $z\sim1$ Type~Ia supernova on the rise. Unfortunately, the transient was outside the footprint of the cadenced WFC3 IR observations. In order to sample the rest-frame optical color of the SN, we requested one orbit of WFC3-IR observations, split between F105W, F125W and F160W. The detection image, together with a larger view of Abell~383, is shown in Fig.~\ref{fig:abell383}.

This cluster was observed Nov 1st 2010 using the  FOcal Reducer and low dispersion Spectrograph 2 \citep[FORS2;][]{1998Msngr..94....1A} for the 8.2 m Very Large Telescope array (VLT) Unit Telescope 1 at Cerro Paranal as part of a spectroscopic follow-up of lensed sources in this field (PID: 086.B-06063(A), PI: Richard). The SN host galaxy was observed for 40 min using the G300V grism and the GG435 order filter, covering the wavelength range 4400-8800 Angstroms. The spectrum shows continuum and a strong emission line identified as [OII] at $z=1.144$, a redshift consistent with the SN color.

As no supernova spectrum was obtained, we must type it using only the photometric data. We follow a procedure similar to that of \citet{2013ApJ...768..166J}. Fortunately, our lightcurve has a well-constrained rise and decline, and \new{measurements in several filters near maximum}.
To represent SN~Ia we synthesize photometry from the template of \citet{2007ApJ...663.1187H} and for non-Ia we use the 51 non-Ia templates (31 SN~II, 20 SN~Ibc) from SNANA \citep{2009PASP..121.1028K}. Each template is fit to  our data and a $\chi^2$ computed. 
\ny{The Core Collapse (CC) supernova templates themselves may be reddened due to dust, and therefore in performing our typing we allow the relative extinction, $\Delta A_V$ to range over both positive and negative values. This distribution of $\Delta A_V$ is likely concentrated around zero, but to be conservative we use a flat prior. To account for the relative reddening we use a Cardelli law \citep{cardelli89}, with $R_V = 3.1 \pm 0.5$, to warp the templates to match the data.}
Also, as the CC templates do not span the full observed range of CC SNe, we add 0.15 magnitudes in quadrature to the error bar on each photometric measurement \new{\citep[for further discussion on these choices, see Appendix in][]{2013ApJ...768..166J}}.
\ny{To be consistent, the same quadrature addition to the photometric error is made for all fits, which will lead to artificially low $\chi^2/dof$ for good SN~Ia matches.}
 For typing purposes, we use the data from ACS F606W to WFC3 F160W, representing the near UV to $i$-band rest-frame.

In \citet{2013ApJ...763...35R}, we considered both how well each individual template matches the data as well as the probability weighting of such templates that do.
The former is a commonly used approach,
while the latter is most appropriate when seeking the correct ensemble statistics (that is, when we wish to know the odds that these particular SNe are of Type~Ia).

We find that the SN~Ia template, compared with the best non-Ia template (SDSS-000018), provides a significantly better fit, with $\Delta \chi^2=7$, indicating that a SN~Ia is preferred at greater than 99\% confidence.
Moreover, the non-Ia template requires $\Delta A_V=-1.0$, i.e. the template is much redder than SN-A1.
(The absolute $\chi^2$ values are 12 and 19 respectively, for 19 dof, with the low $chi^2$/dof caused by the added 0.15 magnitude scatter, as discussed above.)

\ny{Next we examine the probability-weighted fraction of matching templates.
This then needs to be multiplied with the relative observed incidence of observing different SN types. As shown by \citet{2013ApJ...763...35R} the large rate of CC SNe is offset by their faintness, making the probability of finding a Ia and a CC SN close to unity at high redshifts. 
For each template, we compute the relative $\chi^2$ between that template and the best fit. After converting those values into probabilities we can compute the average SN~Ia probability ($=1$, as this is the best fit) and the average CC probability ($=4 \times 10^{-4}$). 
The resulting probability of a SN~Ia relative to the incidence-weighted CC probability is over 99.9\%. The conclusion from this approach agrees with the result using the best-fit SN templates, but in other circumstances these approaches may differ.}

\subsection{SN-H1 -- MACSJ1532.9+3021}

\begin{figure}
\includegraphics[width = 3.in]{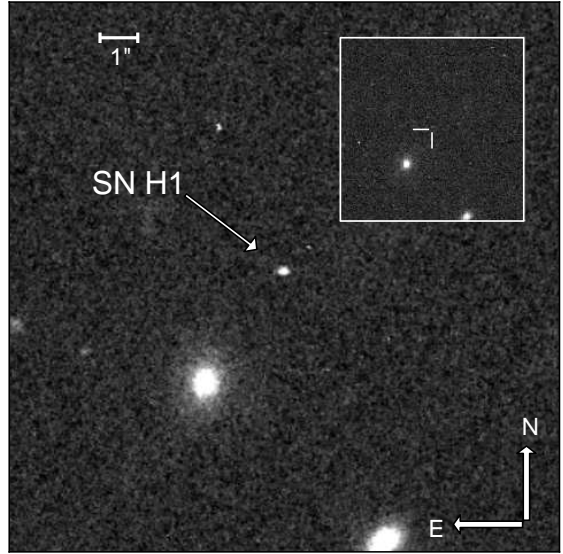}
\caption{SN-H1 behind MACSJ1532; the inset shows the field prior to explosion.}
\label{fig:macsj1532}
\end{figure}

SN-H1 was detected in the field of MACSJ1532.9+3021 \refreply{\citep[MACSJ1532;][]{1998MNRAS.301..881E}}, at $z=0.345$, with coordinates $RA=233.24682$ $Dec=30.36191$ (J2000) in ACS-F625W and F850LP observations taken on March 4 2012 (see Figure~\ref{fig:macsj1532}). 
The scheduled HST observations provided a well-sampled lightcurve with good color coverage, so no additional HST observations were requested.

\begin{figure}
\includegraphics[width = 3.in]{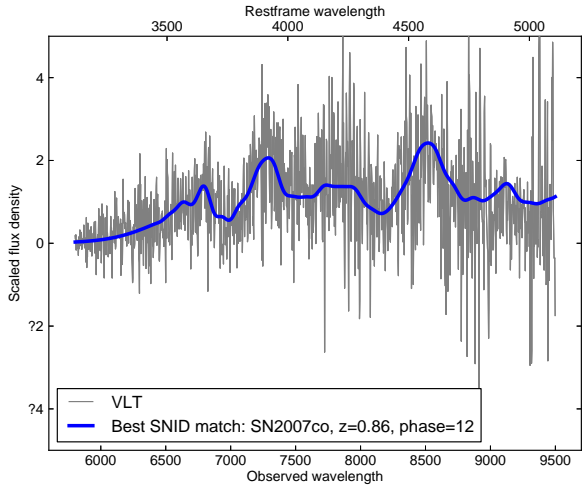}
\caption{VLT observations of SN-H1 together with best SNID match \todo{(Bin spectrum)}.}
\label{fig:vltspec}
\end{figure}

Target-of-opportunity (ToO) long-slit spectroscopy of SN-H1 was obtained from two observatories:
The first, using the Low Resolution Imaging Spectrometer \citep[LRIS;][]{1995PASP..107..375O,2010SPIE.7735E..26R}
optical spectrograph mounted on the 10-m Keck-I telescope at the summit of Mauna Kea with an exposure time of $3\times1000$~sec on March 16th 2012 (600/4000 grism, 400/8500 grating and d560 dichroic; Program ID U043, PI Perlmutter) with seeing $\sim1$ arcsec did not yield sufficient signal-to-noise for conclusive typing and is not considered further.
Fortunately, a ToO the following night at VLT, in $\sim0.7$~arcsec seeing, was successful in yielding a conclusive SN type. A FORS2 spectrum with a exposure time of $7\times1000$~sec was obtained on March 17th 2012 (300I grism, OG590 filter; Program ID 088.A-066, PI Amanullah). 
The Supernova Identification software \citep[SNID;][]{2007ApJ...666.1024B}, applied to the VLT spectrum, securely identifies the transient as a SN~Ia at $z=0.855 \pm0.010$ (See Fig.~\ref{fig:vltspec}). The best match is provided by SN2007co at phase $\sim12$ days past lightcurve maximum, which agrees quite well with SN-H1 lightcurve phase at this time ($\sim10$ days), \new{given typical uncertainties of approximately $\pm 2$ days for spectroscopic dating}.
The SNID \texttt{rlap} parameter is $10.4$ (corresponding to a very strong identification).

\refreply{Type Ibc SNe close to lightcurve peak can exhibit a similar optical spectrum as SNe Ia at phase $\sim 10$. The best non-Ia SNID fit is the peculiar Ibc SN2005bf at phase $-3$, with a significantly worse \texttt{rlap} ($6.3$). We conclude that using only spectroscopic evidence, SNID strongly prefers a SNIa identification for SN-H1. By adding lightcurve phase constraints we can rule out non-Ia SN subtypes.}

\subsection{SN-L2 -- MACSJ1720.2+3536}

\begin{figure}
\includegraphics[width = 3.in]{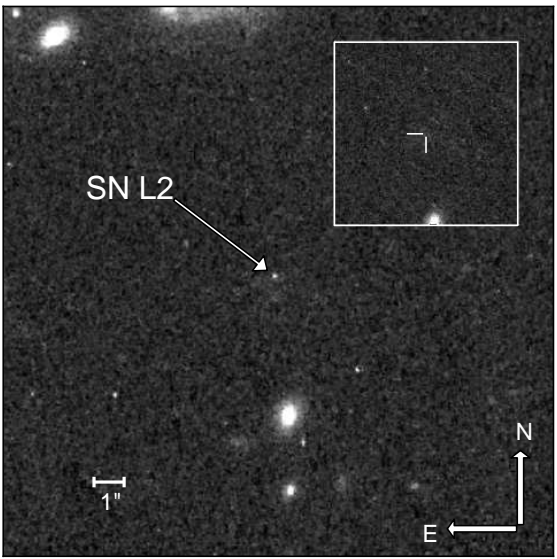}
\caption{SN-L2 behind MACSJ1720; inset shows the field prior to explosion.}
\label{fig:macsj1720}
\end{figure}

Observations of MACSJ1720.2+3536 \refreply{\citep[MACSJ1720;][]{2010MNRAS.407...83E}}, at $z=0.389$, in F850LP on June 17th 2012 revealed two transients: SN-L1 at RA 260.07796 Dec 35.62296 and SN-L2 at RA 260.08757 Dec 35.61133 (Fig.~\ref{fig:macsj1720}). SN-L1 was found in the outskirts of a cluster member galaxy, with photometry compatible with a SN~Ia on the rise in the cluster (SN-L1 is later securely classified as a core-collapse event). SN-L2, on the other hand, had a fainter host for which photometric redshift estimates yielded $1.2<z<1.8$, and a magnitude roughly compatible with an amplified background SN~Ia.

\begin{figure*}
\includegraphics[width = 0.75\textwidth]{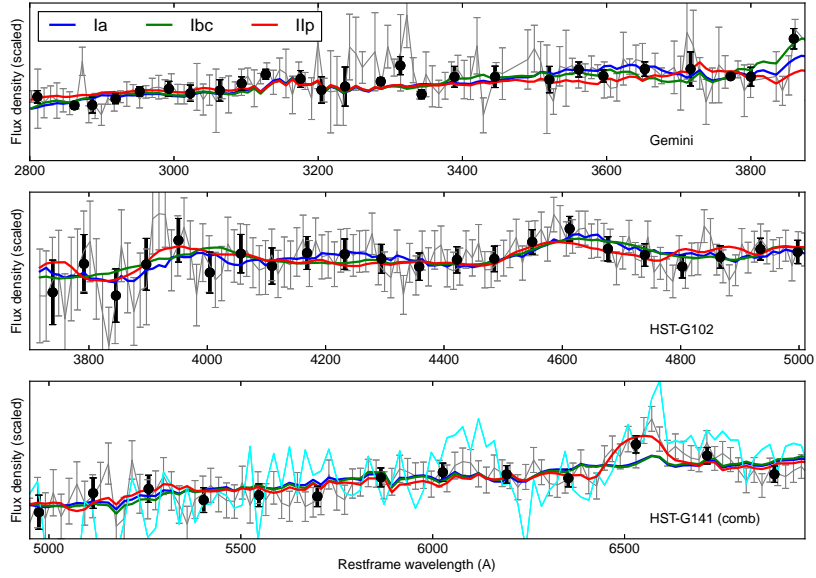}
\caption{\neu{Spectroscopic observations of SN-L2 obtained at Gemini around lightcurve peak (top panel) and with HST $\sim10$ restframe days later using the G102 (mid panel) and G141 grisms (bottom panel). Data is grey, with binned values shown black. We do a combined fit for SN and galaxy template, each with its separate reddening. We fit the fraction of SN light and an offset for each observation. The best fit SN~Ia (blue), SN~Ibc (green) and SN~IIp (red) templates are shown for each spectrum. The lower panel also includes an extraction made on the opposite side of the host galaxy, where no SN light is expected (cyan line).}
}
\label{fig:lgrism}
\end{figure*}

ToO spectroscopic observations, with the slit aligned through both candidates, were made June 30th 2012 with the Gemini Multi-Object Spectrographs \citep[GMOS;][]{2004PASP..116..425H} in longslit mode on the 8.1 m Gemini North telescope at the summit of Mauna Kea with a total exposure time of $1800$~sec (GG455 filter, R400 grating; Program ID GN-2012A-Q-19, PI Perlmutter). 
\new{Both candidates were extracted using the Gemini IRAF GMOS pipeline\footnote{http://www.gemini.edu/sciops/data-and-results/processing-software}. SN-L1 is confirmed as a cluster SN, and as we here focus on lensed SNe this object will not be discussed further. The GMOS spectrum of SN-L2 has low signal-to-noise (see inset in Fig.~\ref{fig:lgrism}), and thus alone can neither confirm nor rule out a high redshift SN~Ia.}

\new{For SN-L2, HST grism observations were then obtained} using both WFC3 G102 ($2200 s$; $R\sim210$; $0.8$ -- $1.15 \mu m$ ) and G141 ($4700 s$; $R \sim 130$; $1.1$ -- $1.7 \mu m$ ) and reduced using the aXe software (Fig.~\ref{fig:lgrism}). One further epoch of G102 observations is not used due to contamination.
\new{We detect $H {\alpha}$ (and low signal-to-noise $H {\beta}$) emission}, allowing us to determine the redshift as $z=1.266\pm0.006$, in good agreement with the photometric redshift estimate.

\neu{
To determine the SN subtype all non-contaminated spectroscopic data (Gemini, HST-G102, HST-G141 orientation 1 and 2) were simultaneously fit with a combination of supernova and host galaxy templates. As supernova templates we use the SALT 2-2 spectral surface \citep{2007A&A...466...11G}, the SN templates compiled by P. Nugent\footnote{http://supernova.lbl.gov/~nugent/nugent\_templates.html} as well as the best fit SNID spectrum of each SN subtype. The exception is the UV spectrum at peak covered by the Gemini observations, \new{which is always fit by one of the Nugent templates since few other spectra extend sufficiently blue}. 
For SNe~Ia, we apply Milky Way type reddening \citep[$R_V=3.1$,][]{cardelli89} according to the color predicted by the SALT2-1 lightcurve fit (see section 4). For other subtypes we fit for the best $A_V$ (allowing negative values). The host galaxy component is best fit with an Sb-like template with $E(B-V)=0.5$ for all supernova templates.
The SN~Ia SN2003it, at phase $+9$ (close to the value predicted by the lightcurve), provides the best fit of the SN~Ia templates ($\chi^2=367$, $dof=333$). The SN~Ibc template fit is as good,  $\chi^2=367$, but for $\Delta A_V \sim -0.6$ (bluer than every known SN~Ibc). The SN~IIp template has worse combined $\chi^2$ ($389$), but is the only template that matches the $H\alpha$ feature well (as this is lacking in the Sb template).
To investigate whether this originates from the SN or the host we extracted the spectrum from the other side of the galaxy, having the same separation from the host core. In this spectrum we find $H\alpha$ that is comparably strong, therefore we believe it is likely that much of the $H\alpha$ in the SN+host spectrum arises from the host.
We conclude that the spectroscopic identification favors SN-L2 as a SN~Ia, but is still ambiguous
(see Fig.~\ref{fig:lgrism}).
}

We turn now to the two photometric classification techniques discussed earlier. We begin with the method based on the best individual matches, and find that with a standard SN~Ia template \citep{2007ApJ...663.1187H} we get a $\chi^2$ of 17.9 for 16 dof.
As previously we allow negative $\Delta A_V$, which allows three CC SNe to fit with $\Delta \chi^2<4$ (but with $-0.8<\Delta A_v<-1.2$). 
The consistent red color of these three SNe may imply that SN-L2, if a CC SN, would have to be much bluer than the current CC sample. For example, we make the a posterori calculation that for equal probabilities of the SN being extincted more or less than SN-L2, the probability of finding all three on the red side is only $2^{-3}$.
Conservatively ignoring this factor, the resulting $\Delta \chi^2$ comparison based on best-matching templates gives a $33$ \% chance that SN-L2 is a SN~Ia.
 
\ny{We now turn to the second method, examining the probability-weighted fraction of matching templates, which is more appropriate for the classification question.
For each template, we compute the relative $\chi^2$ between that template and the best fit. After converting those values into probabilities we can compute the average SN~Ia probability ($=0.526$) and the average CC probability ($=0.03$). 
The resulting probability of a SN~Ia relative to the incidence-weighted CC probability is 95\%. This demonstrates the difference and importance of considering the incidence of comparison objects.} 
\neu{We consider, based on the spectroscopic and photometric evidence, SN-L2 to be a probable, but not certain, SN~Ia.}

\section{Lightcurves and Hubble residuals}\label{sec:data}

\new{The \union analysis of \citet{union2.1} provides a framework for propagating SALT light-curve fits into distances and cosmological constraints. For the lightcurve fits presented here, we take the portion of the framework that computes the sensitivity of the lightcurve fit to each calibration systematic. We also use this framework to compute the $x_1$, $c$, and host-mass correction coefficients. For the host masses, we used Z-PEG \citep{borgne02} on the results of aperture photometry with a 2\arcsec \xspace radius. Note that the host photometry must be de-magnified before a mass can be estimated. 
}

The reduction of the WFC3-IR data, not part of \union, closely follows our previous HST Near Infrared Camera and Multi-Object Spectrometer (NICMOS) reductions. We here give the WFC3-IR specific calibration results, and also discuss how uncertainties were handled for this small set of objects.

\subsection{WFC3 IR Photometry}

In {\union}, we opted to use Point Spread Function (PSF) photometry to extract the NICMOS fluxes, avoiding any resampling of these undersampled images. As the IR imager of WFC3 is significantly more undersampled, we continued with this method. We multiplied each calibrated flat fielded image by the WFC3 IR pixel area map\footnote{\url{http://www.stsci.edu/hst/wfc3/pam/pixel_area_maps}, page updated 09/17/2009} before computing the photometry.

Comparisons between aperture and PSF photometry of the standard star P330E show that the TinyTim \citep{1995ASPC...77..349K} PSF is systematically too narrow, causing the flux derived from the PSF photometry to be \psfphotlowisolated below that from the aperture photometry. We thus fit for a convolution kernel that matched TinyTim PSFs to \new{HST calibration observations of P330E}. The convolution kernel was allowed to vary radially, but was constrained to have elliptical symmetry. 
\ny{In constructing this PSF we were careful to simulate the conditions when measuring supernova fluxes. Because supernovae are faint, the background dominates the noise and therefore PSF fitting weights each pixel nearly equally. We thus} assume equal uncertainties per pixel, while simulating a fit of host galaxy light.

The PSFs generated with this approach followed the data well; the new PSF photometry matched \new{aperture} photometry to less than \psfphotmatch on average for all filters. 
Checking individual PSF photometry measurements against aperture photometry shows a residual 0.02 magnitude \ny{scatter, representing focus variation and small variations in the PSF with position. We add this scatter in quadrature to the statistical uncertainties}. This uncertainty is also appropriate for ACS photometry.

Using our PSFs on data for P330E (again assuming that all pixels have equal weight, similar to SNe), we find zeropoints~{$\sim0.02$} mag fainter than the STScI zeropoints\footnote{\url{http://www.stsci.edu/hst/wfc3/phot_zp_lbn}, page updated 03/06/2012}. For F105W, F110W, F125W, F140W, and F160W, we find {\brightzps} \new{on the VEGAmag system}. Subtracting 0.03~magnitudes for the count-rate non-linearity (discussed more below), gives the zeropoints we used in our analysis, \faintzps.

As with some of the models used in \citet{union2.1}, we modeled the host galaxy in each WFC3 filter with a two-dimensional second-order spline plus a PSF for the supernova. The relative alignment of each image was included in the fit, as was residual variation in the sky level. The photometry was stable to reasonable changes in the spline \ny{node} spacing. For the data in each filter, we placed simulated supernovae on the host galaxy at positions with similar amounts of host galaxy light to verify our parametrization of the host galaxy. For SN-A1, which lacks reference images, we used a spline node spacing of 0.36~arcsec (just under three pixels). For SN-L2, which has reference images, we used 0.144~arcsec, or just over one pixel. SN-H1 does not seem to have structured underlying galaxy light, so it made no difference for the WFC3-IR data if we modeled it (for the results presented here, we used a node spacing of 0.72~arcsec).

\subsection{SALT lightcurves}

\begin{figure*}
\includegraphics[width = 2.5in]{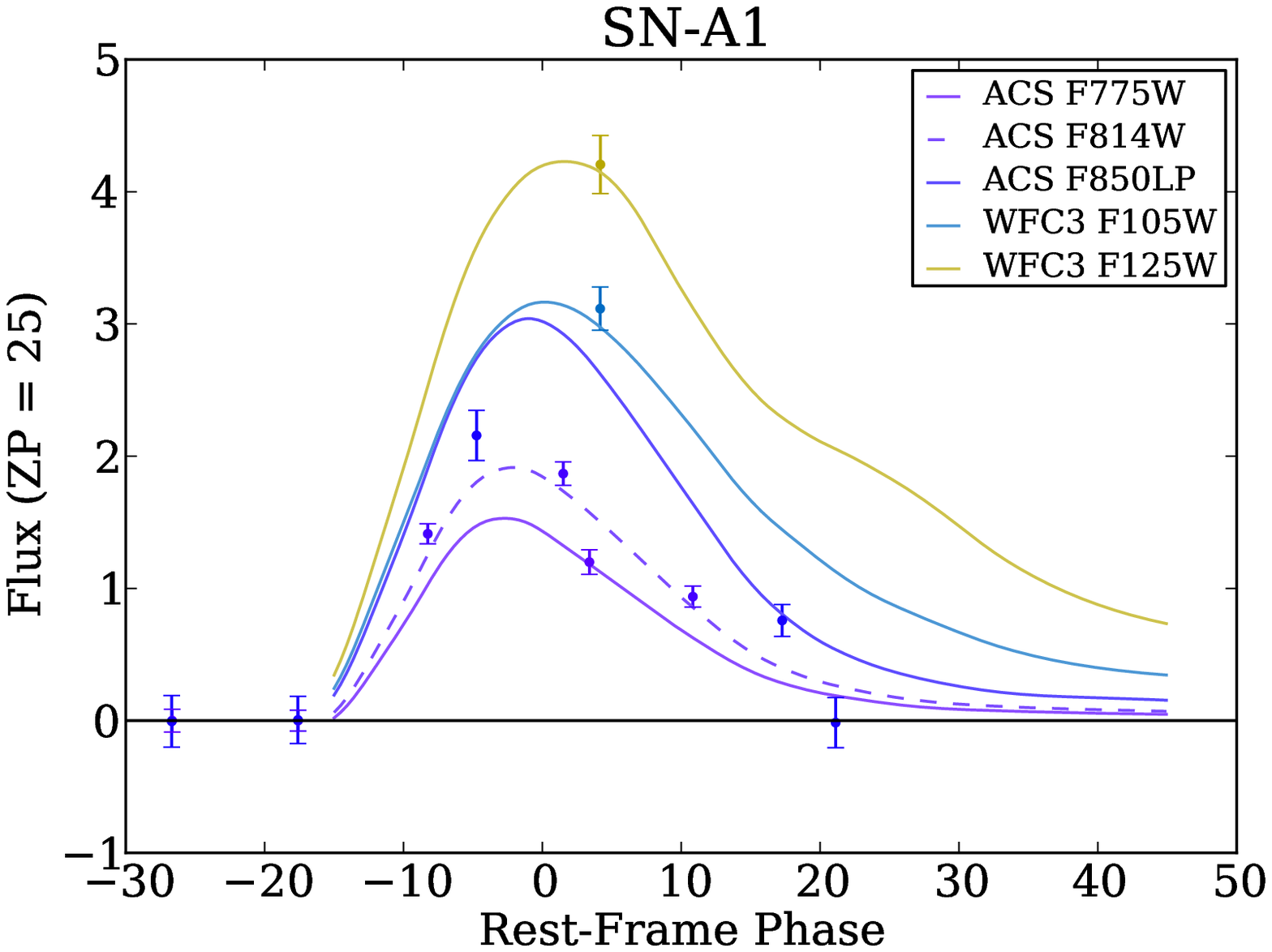}
\includegraphics[width = 2.5in]{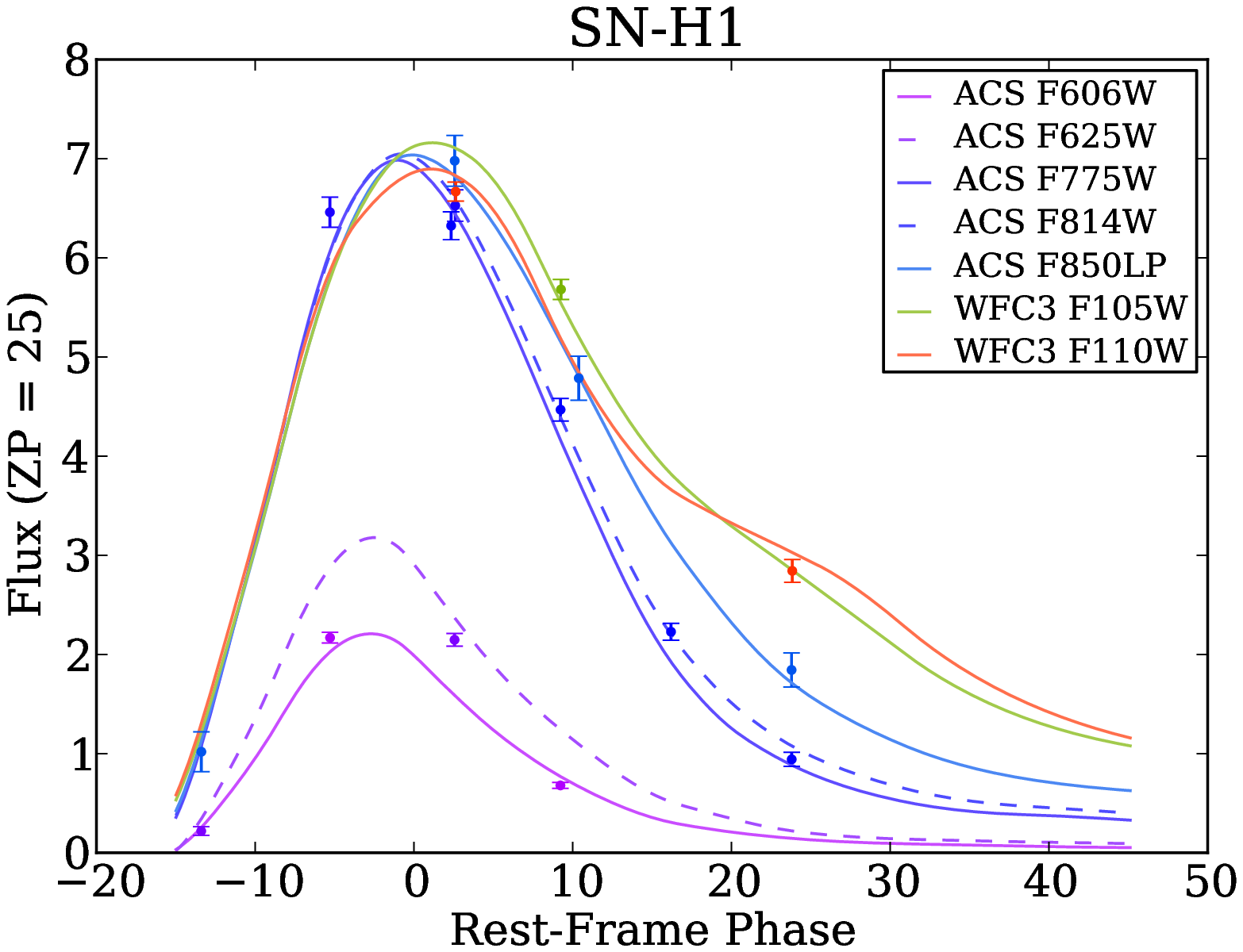}
\includegraphics[width = 2.5in]{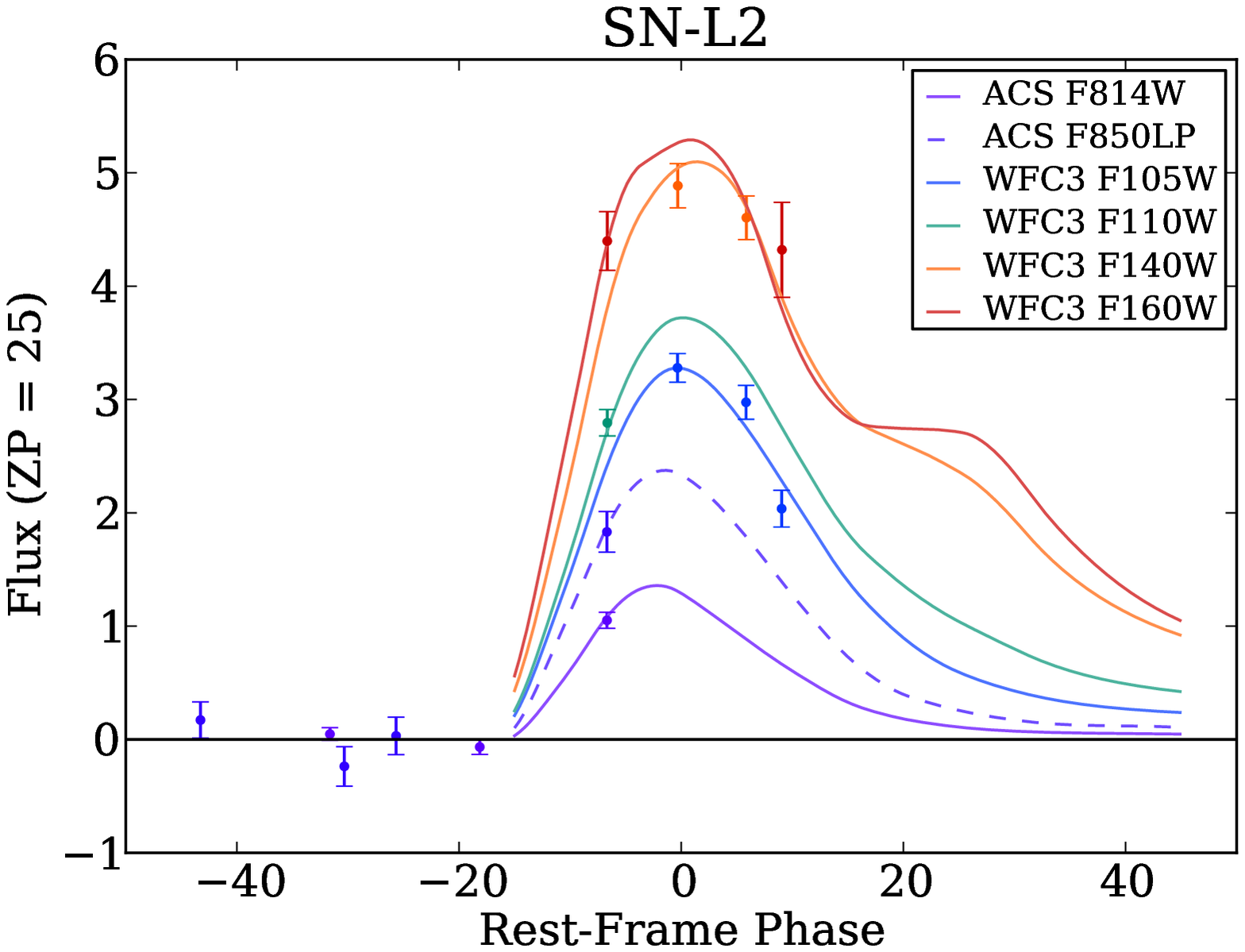}
\caption{Lightcurves of: SN-A1, SN-H1 and SN-L2 (left to right). For plotting purposes, we arbitrarily subtract the weighted mean of the underlying galaxy light for each ACS band. When fitting lightcurves in the analysis, the covariance due to unknown underlying galaxy flux in each band is also included.}
\label{fig:lc}
\end{figure*}

Light-curve fits were initially made using the SALT2-1 light-curve model. The improved SALT2-2 is currently available, but we had decided to use SALT2-1 before the blinding was lifted. As will be discussed below, the amplification estimate of SN-A1 varies significantly depending on which model version is used. 
Changes for SN-H1 and SN-L2 are negligible.
We take the light-curve shape and color correction coefficients, the mass-correction coefficient, and the absolute magnitude ($h=0.7$) from \citet{union2.1}: $\alpha = 0.13$, $\beta = 2.47$, $\delta = -0.03$, and $M_B = -19.32$. (Later when we use SALT2-2, we will use the values from Rubin et al. (in prep): $\alpha = 0.14$, $\beta = 3.07$, $\delta = -0.07$, and $M_B = -19.09$; \refreply{the change in the fiducial absolute magnitude, $M_B$, is mostly due to an arbitrary redefinition of the color zeropoint.}) 
The SALT2-1 light-curve fits are shown in Figure~\ref{fig:lc} and the parameters are provided in Table~\ref{table:salt}.

\subsection{Statistical Uncertainties}

The following sources of statistical uncertainty were included, following the {\union} analysis \citep{union2.1}: light-curve parameter uncertainties, {\snia} \new{intrinsic dispersion} ($0.11$~mag), and 16\% uncertainty in the MW extinction map from \citep{sfd98}. \new{The intrinsic dispersion value is taken from near-IR-observed HST SNe. Note that when performing cosmological analysis our error bar would include uncertainty due to gravitational lensing. However, in this context, lensing is our signal and is therefore not included in the statistical error budget.} 

\begin{savenotes}
\begin{table*}
\begin{tabular}{l| c| c| c| c| c| c | c}
SN & z & $m_B$ & $x_1$ & $c $ & Host Galaxy Stellar Mass\footnote{In units of $log(M/M_{\sun})$} &$\Delta m_{SN}$\footnote{These values are statistical uncertainties only, and do not include the conservative correlated 0.05 magnitude uncertainty described in Table \ref{table:sys}. When computing the uncertainty on the ensemble mean, we do include the correlated uncertainty.} & $\Delta m_{map}$ \\
\hline
\hline
A1 & $1.144 \pm 0.005$ & $25.23 \pm 0.04$ & $ ~0.62 \pm 0.57$ & $ 0.14 \pm 0.04$ & $10.7 \pm 0.1$ &$\snAmag \pm \snAmagerr$ & \mapAmag \\
H1 & $0.86 \pm 0.01$ & $24.03 \pm 0.03 $ & $ ~0.56 \pm 0.40$ & $-0.07 \pm 0.03$ & $<9.1$ &$\snHmag \pm \snHmagerr$  & \mapHmag\\
L2 & $1.266 \pm 0.006$ & $25.35 \pm 0.05 $ & $-0.21 \pm 0.83 $ & $0.26\pm 0.05 $ & $10.9 \pm 0.2$ & $\snLmag \pm \snLmagerr$ & \mapLmag
\end{tabular}
\caption{SALT2-1 lightcurve parameters and predicted magnification from SN distances ($\Delta m_{SN}$) and lensing maps ($\Delta m_{map}$). \new{$m_B$ is the peak $B$ band magnitude, $x_1$  measures the lightcurve width and $c$ the lightcurve color. For all SNe, the difference between $\Delta m_{SN}$ and $\Delta m_{map}$ can be compared with the measured intrinsic dispersion, $0.11$, of \snia with similar data in \union.
\label{table:salt}
}}
\end{table*}

\subsection{Systematic Uncertainties}

We follow our \union analysis for the systematic uncertainties, but do remark on new WFC3-specific uncertainties. A summary of their impact on the distance modulii is given in Table~\ref{table:sys}. 
As we are working with a small number of SNe, the combined uncertainties will be dominated by statistical uncertainties. We can therefore make a highly conservative systematics analysis, and we note that these systematics can be substantially reduced in the future. 
\end{savenotes}

\citet{2010wfc..rept....7R,riess11} finds that the WFC3 IR detector exhibits a small ($\sim 0.01$ magnitudes per dex) count-rate non-linearity. Although there is no official non-linearity correction code available, we follow their recommendations and correct our zeropoints brighter by $0.03 \pm 0.01$ magnitudes, with the uncertainty correlated across all WFC3 filters.

Another possible source of non-linearity is variation in the interpixel capacitance with counts. Results from \citet{hilbert11} indicate that there could be an effect as large as 0.01 magnitudes when comparing our supernova photometry to the much-brighter standard stars. As with the count-rate non-linearity uncertainty, we assume this 0.01 magnitude uncertainty is correlated. Comparison of PSFs at different flux levels would calibrate out this systematic, but this is not necessary for our analysis.

As noted above, P330E was the source of our WFC3 PSFs, and we therefore account for systematic difference due to the Spectral Energy Distribution (SED) difference between P330E and our SNe. Redoing the PSF photometry with PSFs from a range of filters reveals that the photometry changes $\sim 0.05$ magnitudes per 1000~\AA \xspace change in effective wavelength. P330E should match our SNe in effective wavelength to within $\sim 200$~\AA \xspace for most filters, or to within $\sim 400$\AA \xspace for the broad F110W. We thus add a 0.02 magnitude correlated uncertainty on the F110W photometry, and a 0.01 magnitude correlated uncertainty on the other WFC3 IR photometry. Careful modeling of stars with differing colors can greatly reduce this systematic, but we do not need to attempt that here.

As noted above, we find WFC3 IR zeropoints $\sim 0.02$ magnitudes fainter than the STScI values. It is possible that this is a difference in encircled energy normalization, but to be conservative and since the effect is small compared to the amplifications we wish to measure, we take a 0.02 magnitude uncertainty on each zeropoint. 

Finally, we use a background cosmology of flat $\Lambda$CDM, with $\Omega_m = 0.30 \pm 0.02$, which gives an (essentially correlated) uncertainty of about 0.026 magnitudes for our SNe. We also take a 0.03 magnitude uncertainty on the absolute magnitude \new{\citep[dominated by calibration, see][]{union2.1}}. 
\new{These last two effects make up the bulk of the correlated uncertainty in this analysis.}
\ny{When summed using the covariance matrix, these effects are $0.05$ mag in total. This is comparable to the expected systematic error in cluster mass reconstructions, but much less than the uncertainty on individual standardized SN brightnesses.}

\begin{table}
\begin{tabular}{l| r}
SN Lightcurve Systematics & Magnitudes\\
\hline
\hline
ACS and WFC3 Zeropoints & $-0.02$ to 0.03\\
Near IR Flux Reference & 0.02\\
WFC3 IR Count + Count-Rate Non-Linearities & 0.02\\
Uncertainty in WFC3 PSFs & $-0.01$ to 0.03 \\
Uncertainty in Distance Modulus & 0.03 \\
Uncertainty on Absolute Magnitude & 0.03 \\
Other Systematics from \union & 0.02 \\
\hline
Total, summed in distance modulus covariance matrix & 0.05 \\
\end{tabular}
\caption{
Sources of systematic uncertainty, following {\union} \citep{union2.1}. The systematics that are new to WFC3 data and this analysis are broken out. The typical effect of each systematic uncertainty category on the distance modulii is given. Negative systematic uncertainties indicate anti-correlation between our SNe, caused by the range of redshifts (e.g., increasing the ACS F850LP zeropoint makes SN-H1 bluer, but SN-A1 redder).
\label{table:sys}
}
\end{table}

\section{Magnification predictions from cluster mass models}\label{sec:maps}

\subsection{Procedure}

For each of the three CLASH clusters, we have constructed parametric models of the mass distribution based on constraints from the strong lensing observed in the cluster cores. The model parameters have been 
optimized with Lenstool\footnote{available at http://projets.lam.fr/projects/lenstool/wiki}
\citep{2007NJPh....9..447J,2009MNRAS.395.1319J} using a Bayesian Markov Chain Monte-Carlo (MCMC) sampler. Based on a sample of $\sim100$ models sampling the posterior probability-density function of all parameters, we can predict the 
average magnification and statistical error (under the assumptions of the parametric models) at the 
locations of the supernovae. The procedure we use is very similar to previous published work on cluster 
cores \citep[e.g.][]{2007ApJ...668..643L,2009A&A...498...37R, 2010MNRAS.404..325R}.

\begin{figure*}
\includegraphics[width = 2.3in]{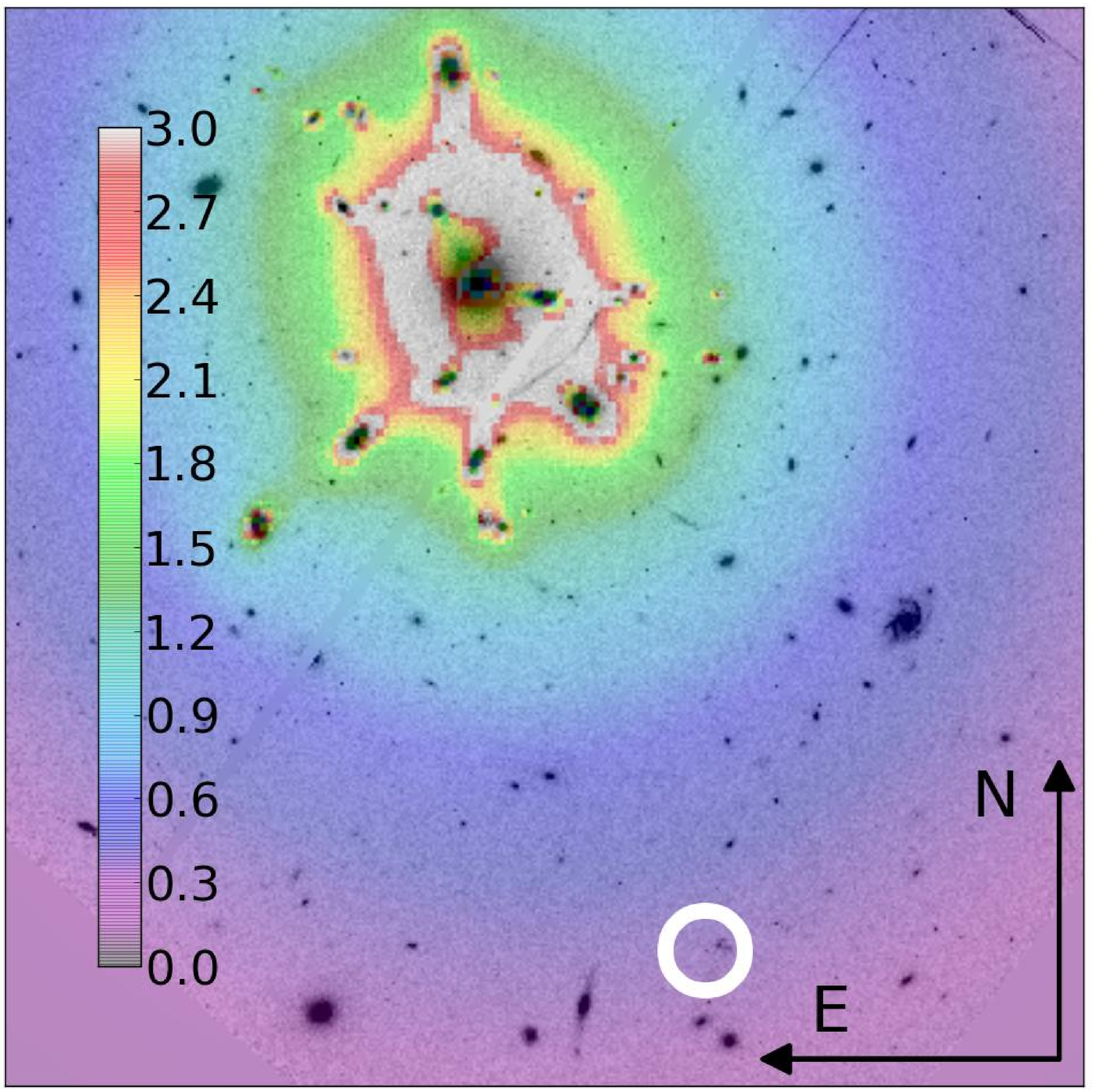}
\includegraphics[width = 2.3in]{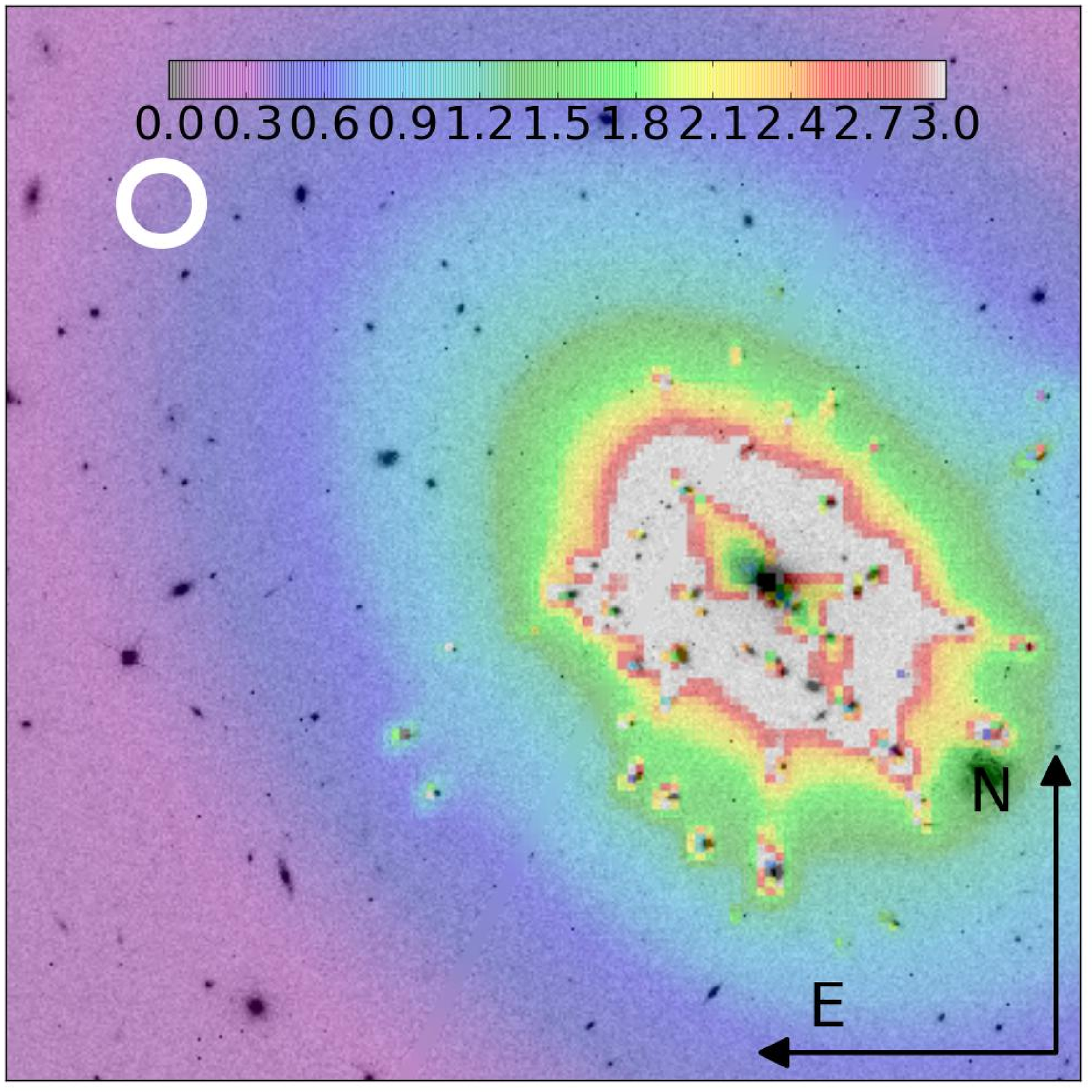}
\includegraphics[width = 2.3in]{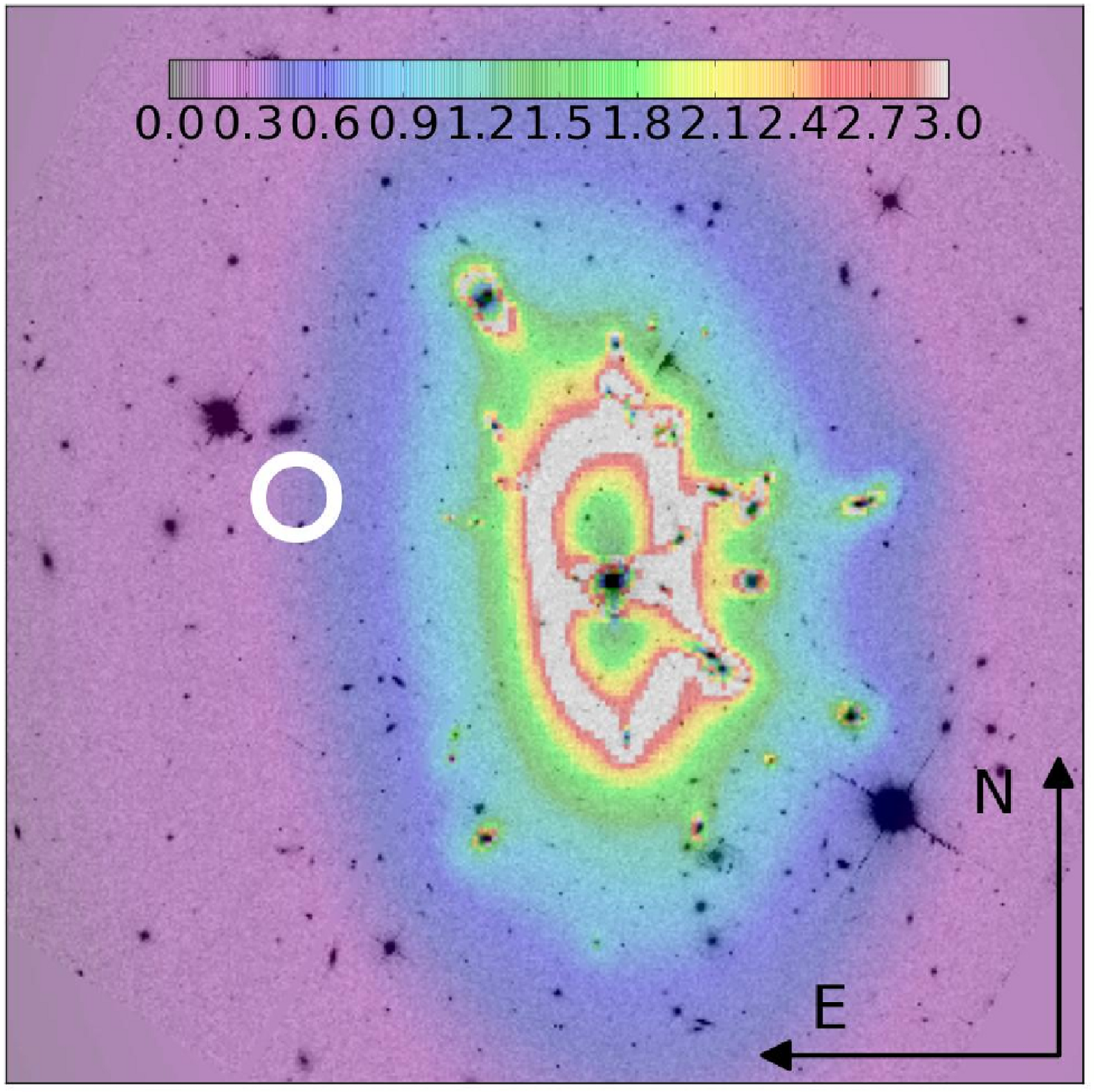}
\caption{Magnification models for, from left to right: Abell 383, MACSJ1532, MACSJ1720. The SN position is marked with a white circle. \new{The colorbar shows predicted brightness amplification, expressed in magnitudes.}
}
\label{fig:maps}
\end{figure*}

Full details on the modeling of each cluster and the resulting mass distributions have either been presented in 
\citet{2011MNRAS.414L..31R} (for Abell 383) or will be published in a forthcoming paper (Richard et al. 2013, 
in preparation), but we summarize the main ingredients of each model in the following subsections.
In addition, since all three supernovae from our study are located at larger clustercentric distance than the 
strong lensing region, the error on the magnification will be dominated by the systematic error on the 
assumed cluster mass profile, which is typically truncated at $\sim1$ Mpc (see \citet{2007ApJ...668..643L} and  
\citet{2010MNRAS.402L..44R} for a discussion). In order to better estimate this additional source of error, we recomputed the magnification letting the truncation radius vary between 500 kpc and 2 Mpc. Figure~\ref{fig:maps} shows magnification contours for the three clusters, and the magnification estimates are collected in Table~\ref{table:salt}.

\subsection{Abell 383}

The cluster mass distribution is constrained by the location of six multiply-imaged systems, five of which have been confirmed with spectroscopy \citep{2013ApJ...765...24N}. At the location of supernova SN-A1 and for a redshift $z=1.144$, our Lenstool mass model predicts a magnification of 1.40$\pm0.02$ (linear value, statistical error from MCMC samples). By varying the cut-off radius of the mass distribution we estimate the systematic error to be $\pm0.07$. In total, the magnification is estimated to be \mapAmag mags.

\subsection{MACSJ1532}

The cluster mass distribution is constrained by the location of only one multiple system with a spectroscopic redshift at $z=0.87$, very close to the redshift of SN-H1. At the location and redshift of SN-H1, we predict a magnification factor of 1.39$\pm0.03$ (statistical error) and estimate a systematic error of $\pm$0.06 by varying the cluster profile. In total, the magnification is estimated to be \mapHmag mags.

\subsection{MACSJ1720}

The cluster mass distribution is constrained by the location of two multiple systems, one of which has a clear photometric redshift at $z=0.7\pm0.1$ \citep[based on the public CLASH photometric redshift catalogs;][]{2012ApJS..199...25P}.
We created a variety of mass models by varying the redshift constraint on this multiple system in the range $0.6<z<0.8$ and derive the magnification factor 1.42$\pm$0.09 (linear value and statistical error) at the location and redshift of SN L2. Again, by varying the mass profile on these models we estimate a systematic error of $\pm$0.06. In total, the magnification is estimated to be \mapLmag mags.

\emph{After unblinding}, \new{we identified a likely counterimage for the main multiple system used in our strong lensing model. Adding this new constraint shifts the estimate up to  1.65+/-0.12 (combined).
Further, including a foreground ($z \sim 0.2$) galaxy located near the supernova will potentially enhance the magnification to $1.71 \pm 0.12$, or $-0.58 \pm 0.08$ mags.}

\section{Discussion}\label{sec:disc}

All candidates show $>1 \sigma$ differences between mass model and supernova prediction after unblinding (see Table~\ref{table:salt}). 
\new{Seeing such large dispersion in all three candidates is very unlikely, and we will therefore} examine each candidate separately. We will see that this large dispersion can all be accounted for.
This finding will, in turn, lead to a discussion about the importance and methods of blinded analysis.

\subsection{SN-A1 --- Abell 383}

While the restframe optical spectra of SNe~Ia are very well studied, only a handful of nearby SNe have high signal-to-noise observations covering UV wavelengths \citep[see e.g.][]{2012MNRAS.426.2359M}. This is further complicated by changes with progenitor metallicity, which are thought to be much greater at bluer wavelengths \citep{2008MNRAS.391.1605S,2012MNRAS.427..103W}. The SN-A1 lightcurve is dominated by such UV observations, with restframe optical colors only at one epoch.

This UV template uncertainty is manifest by a change in brightness estimate of as much as $0.2$ magnitudes, depending on which version of SALT (the SN lightcurve fitting tool) is used.
As seen in Table~\ref{table:salt}, using SALT2-1, SN-A1 ends up $0.2$ magnitudes fainter than predicted by the mass model. However, as reported in Table~\ref{table:postunblindingtable} with the updated SALT2-2 model, introduced as this work progressed and thus not our default fit version, the predicted SN brightness excess is $\snAmagSALTtwotwo$ magnitudes, identical with the mass model prediction. This $> 1\sigma$ move (for the same input data) shows that the uncertainties in the rest-frame UV model may have been underestimated in SALT2-1 (the stated uncertainties are larger in SALT2-2).
We note that the brightness estimates of the other SN candidates, which have more rest-frame optical data, do not change with SALT version. Due to the uncertainties in the rest-frame UV, ideally SNe should be observed in rest-frame optical bands with multiple epochs. 
\new{It is possible that \snia UV fluxes are as standardizable as at redder wavelengths, just less well measured and/or modeled \citep[][]{2013arXiv1308.2703M}. Future lightcurve fitters might thus be able to also standardize \snia well in the UV.}
\subsection{SN-H1 --- MACSJ1532.9}

SN-H1, on the other hand, has a very well-measured lightcurve and thereby small measurement errors, and is \snHfainter mag fainter than predicted by the mass map (\snHfaintersigma $\sigma$), consistent with having experienced no amplification. 
With the current data we can not rule out that this measurement corresponds to a statistical fluctuation within the SN intrinsic dispersion.
We note that if we \emph{assume} that there is no structured hostlight underlying SN-H1 and fix the underlying ACS galaxy light to zero, the amplification increases to $-0.24$ magnitudes, within 1 $\sigma$ of the map estimate. However, as we had decided to follow {\union} in using floating band offsets before unblinding, this is clearly a post-unblinding choice, and is therefore listed in Table~\ref{table:postunblindingtable}. 
Whether structured host light is present under SN-H1 can be straightforwardly settled by obtaining deep reference observations after the SN light has faded, but that additional step is not needed in this pilot study.
\footnote{After completing the manuscript we became aware of a potential high-mass host for SN-H1 located at a projected distance of $\sim 30$ kPC \citep{2013arXiv1312.0943P}. This would change the appropriate mass step correction, and thus the amplification, by $3\%$. Our conclusions are unaffected by this small change.}

\begin{savenotes}
\begin{table*}
\begin{tabular}{l| c| c| c| c| c| c | c}
SN & $m_B$ & $x_1$ & $c $  &$\Delta m_{SN}$\footnote{These values are statistical uncertainties only, and do not include the conservative correlated 0.05 magnitude uncertainty described in Table \ref{table:sys}. When computing the uncertainty on the ensemble mean, we do include the correlated uncertainty.} & $\Delta m_{map}$ \\
\hline
\hline
A1 & $25.26 \pm 0.05$ & $ 0.30 \pm 0.73 $ & $ 0.10 \pm 0.05$ &$ -0.38 \pm 0.21 $ & \mapAmag \\
H1 & $ 24.05 \pm 0.02 $ &  $ 0.17 \pm 0.19 $ & $ -0.10 \pm 0.02 $&$ -0.30 \pm 0.13 $  & \mapHmag \\
L2 & $ 25.34 \pm 0.06 $ & $0.27  \pm  0.63$ & $0.16 \pm 0.04$ &  $ -0.75 \pm 0.15 $ & $ -0.58 \pm 0.08 $\\
\end{tabular}
\caption{These values have been updated after unblinding, but represent our current estimates. To summarize, we switch lightcurve fitters from SALT2-1 to SALT2-2, add more structure to the magnification map of SN-L2, and assume that the host-galaxy light underneath SN-H1 is smooth.
\label{table:postunblindingtable}
}
\end{table*}

\subsection{SN-L2 --- MACSJ1720}

Using the magnification map available at the time the analysis was unblinded, SN-L2 deviates in the opposite direction by \snLbrigher magnitudes, or $\snLbrighersigma \sigma$, brighter than predicted by mass maps. 
As discussed in section 5.4, the new strong lens candidate and the massive foreground galaxy that have been introduced in the lensing model increase the magnification map prediction by $\sim 0.13$ magnitudes, with a combined (SN+lens map) uncertainty $0.17$ mags. This makes the deviation $<1 \sigma$.

We note that MACSJ1720 is the only system without spectroscopic confirmation of the strongly lensed system and that the mass model uncertainty is consequently larger. It would be best if future studies were to predefine criteria \new{required to consider a cluster magnification map complete prior to comparison with the accompanying supernova amplification.}

We also note that the classification of SN-L2 as a SN~Ia is considered likely but not secure. See e.g. \citet{2013ApJ...768..166J} for further discussions on the challenge when typing high redshift SNe using HST grism and/or photometric data.

\subsection{Ensemble results}

\neu{We now examine the SN amplification and the predicted clusters mass model magnification predictions for the blinded study as an ensemble using the values of $\Delta m_{SN}$ and $\Delta m_{map}$ given in Table~\ref{table:salt}.  We find an ensemble mean of $\Delta m_{\mu} = 0.09\pm0.09^{stat}\pm0.05^{sys}$~mag and dispersion of $\sigma_{\mu}=0.21$~mag. This dispersion is higher than expected from the SN and lensing map uncertainties, but dispersions of at least this size occur by chance 17\% of the time in such a small sample. Because the sample size is small, rather than using the observed dispersion we have used the uncertainties derived from the quoted uncertainties on the supernova lightcurve measurements and the lensing model amplification when calculating the error in the mean.  Overall, the mean agreement for the ensemble found in the blinded analysis is already quite good despite some of the individual deviations described above being slightly large.

Following the same approach for the results of the post-unblinded analysis, as presented in Table~\ref{table:postunblindingtable}, we find an ensemble mean of $\Delta m_{\mu} = -0.03\pm0.09^{stat}\pm0.05^{sys}$~mag and dispersion of $\sigma_\mu=0.12$~mag. This agreement is excellent, however, we caution against overinterpreting the quality of the agreement since these values result from changes made after unblinding. Nonetheless, the changes that produced this improvement are well motivated.  In the case of the cluster lensing model, a new strong lensing counterpart was identified, and a foreground massive galaxy was added.  In the case of the switch from SALT2-1 to SALT2-2 for the SN analysis, by almost any metric SALT2-2 has been found to perform better in fitting SNe~Ia lightcurves (see Rubin et al, in prep).}

\subsection{Using SNe~Ia as tests of cluster lens maps}

The upcoming HST-Frontier survey\footnote{\texttt{www.stsci.edu/hst/campaigns/frontier-fields/}} aims at providing high precision lensing mass models, which will be used both to study cluster properties and to probe the largely unknown high redshift universe that the magnification allows us to see.
\neu{Already, several different methods for creation of mass models exist. 
Evaluating the model accuracies will be a key element in fully utilizing the new data.}

The SNe detected in this pilot study show that a larger sample of SNe~Ia, with good lightcurve coverage, could be used as ``test beams''. 
Our study highlights the importance of a blinded analysis framework: possible strong lenses or substructure could potentially be added gradually until the results meet expectations, and variations in supernova lightcurve analysis could be tried, in an effort to minimize deviations. Blinded analysis requires a decision of when this process is ``done'' \neu{before looking at the final results. }

\ny{Current models suggest that substructure in dark matter halos is not likely to create magnitude differences beyond $0.05$ magnitudes. To accurately measure such deviations  with SNe, given their current magnitude dispersion, would require $\sim100$ such cases. However, there may be unanticipated scenarios in which a small sample can yield exciting results. 
Furthermore, improvements in SNe~Ia standardization techniques would also improve the sensitivity.}
Several methods for doing exactly this have been demonstrated using nearby SNe~Ia \citep{2009A&A...500L..17B, 2011ApJ...731..120M, 2012MNRAS.425.1007B,2013ApJ...766...84K}.

Unfortunately, obtaining $>20$ amplified SNe will be a challenge. The CLASH survey, though not optimized for transient detections, yielded roughly one lensed SN~Ia per cluster per one year of monitoring. A large scale survey would demand monitoring of at least 10 clusters for one year, with frequent high quality follow-up of all detected supernovae. \new{A smaller set of SNe~Ia, if observed close to the cluster core, could provide interesting limits on any error on the overall magnification scale, due to the much larger magnification expected here. However, the effective volume probed, and thus the detection probability, will drop in proportion.}

\refreply{Alternative ways of using lensed SNe~Ia have been suggested. \citet{2011A&A...536A..94R}, e.g., simulated how lensed SNe can be used as additional constraints when constructing the mass map.}
The method we are suggesting here has the advantage of providing an independent test of strong lensing mass maps in general -- we expect only a small subset of all clusters to host detected background \snia.

\new{Finally, the chance of finding a SN in a strongly lensed background galaxy is small, but only one such object (of any kind) could provide an independent measurement of the Hubble parameter through a measurement of the time-delay \citep{1964MNRAS.128..307R}.} 

\end{savenotes}

\section{Conclusion}\label{sec:conc}

We have presented three SNe Ia detected behind clusters observed as part of the CLASH survey. 
The small peak magnitude uncertainties for SN-H1 and SN-L2 (totally dominated by the SN intrinsic dispersion) are remarkable since these observations were made in a novel way, using a mixed selection of filters with irregular cadence. This further demonstrates HST/WFC3-IR capabilities for precision SN measurements at high redshifts.

The SN luminosities were compared with those predicted from strong gravitational lensing maps. The results of this comparison are as follows:
\begin{enumerate}[leftmargin=0.5cm]
\item In SN-L2, we now have a clear example of a SN~Ia significantly ($\sim5\sigma$) amplified by a foreground galaxy cluster.
\item We find remarkably good agreement between these SNe~Ia and
   the mass models of their clusters, with a difference of $\Delta m_{\mu} = 0.09\pm0.09^{stat}\pm0.05^{sys}$~mag
   from our blinded analysis, and $\Delta m_{\mu} = -0.03\pm0.09^{stat}\pm0.05^{sys}$~mag after additional
   adjustments were made.
\item Substructure would primarily add dispersion and it is thus comforting that we find a dispersion of only $\sigma_{\mu}=0.21$~mag from our blinded analysis, and an impressive $\sigma_{\mu}=0.12$~mag after additional adjustments. 
\end{enumerate}

Such comparisons can in principle be used to test assumptions regarding the properties of dark matter halos, but would need statistical samples significantly larger than what is currently available.

Based on the three SNe in this pilot study we can provide several important guidelines for future larger surveys:
\begin{enumerate}[leftmargin=0.5cm]
\item SN Ia UV flux variations are still not well-understood and therefore multiple rest-frame optical observations are needed for a reliable constraint.
\item \new{Mass models, including analysis of structure along the line of sight, should be completed before amplification comparisons are performed.}
\item An explicit choice should be made and reported as to whether the SNe are used unblinded to improve the model, or blinded to test the model.
\end{enumerate}

With these ideas in mind, there is strong motivation to pursue a larger sample of lensed SNe~Ia, in order to verify cluster mass models, break the mass-sheet degeneracy and potentially probe dark matter properties \neu{or measure the Hubble constant in a new way}.

\section*{Acknowledgments} 

We would like to acknowledge the CLASH team for planning and carrying out the survey that made this analysis possible, and Saurabh Jha and Brandon Patel for stimulating discussions.
We are grateful to Jeffrey Silverman and Brad Cenko, who obtained the Keck ToO observations of SN-H1 (while observing for a PI: Filippenko program). 
HA and JPK acknowledge support from the ERC advanced grant LiDA. JPK also acknowledges support from CNRS. JR is supported by the Marie Curie Career Integration Grant 294074. RA and AG acknowledge support from the Swedish Research Council and the Swedish National Space Board. 
This work supported in part by the HST program GO/DD-12360. 
This work was also partially supported by the Directory, Office of Science, Department of Energy, under grant DE-AC02-05CH11231.
Based in part on observations made with the NASA/ESA Hubble 
  Space Telescope, obtained from the data archive at the Space
  Telescope Institute. STScI is operated by the association of
  Universities for Research in Astronomy, Inc. under the NASA contract
  NAS 5-26555. The observations are associated with program GO/DD-12360. 
ESO-VLT observations made under Program ID 088.A-0663(A) (PI: Perlmutter). Keck observations made under proposal U043. 
\foreignlanguage{portuguese}{
Based on observations obtained at the Gemini Observatory, which is operated by the
Association of Universities for Research in Astronomy, Inc., under a cooperative agreement
with the NSF on behalf of the Gemini partnership: the National Science Foundation (United
States), the National Research Council (Canada), CONICYT (Chile), the Australian Research 
Council (Australia), Ministério da Ciência, Tecnologia e Inovação (Brazil) and
Ministerio de Ciencia, Tecnología e Innovación Productiva (Argentina). Gemini programme ID GN-2012A-Q-19.
}

\bibliographystyle{mn2e}
\bibliography{magnifiedsn_mnras}

\appendix
\section{SN photometry}

Below we present the multiband $HST$ photometry for SN-A1, SN-H1
and SN-L2. For each SN we list the date of observation, both as
calendar dates and modified Julian dates. We then list the filter,
exposure time, measured flux for each observation. Observations of reference images are also listed, with no flux measurement quoted. Next, the diagonal uncertainty,
that is, the portion of the uncertainty that is uncorrelated between
the filters, is given. To aid the reader in converting fluxes to
magnitudes, we provide the zeropoint in each filter on the VEGAmag
system. Here the values used for WFC3 are those determined by us
in Section 4.1. The off-diagonal values of the covariance matrix
are then listed; these arise from our method of accounting for underlying light from the host. The last column lists the
$HST$ program identification numbers: GO-12065, GO-12454, GO-12455 PI:~Postman, GO-12099
PI:~Riess, and GO-12360 PI:~Perlmutter.

\begin{table*}
\begin{tabular}{l  l   l   l  r  r  r  r  r}
\hline

UT Date & MJD & Filter & Exp. Time & Flux & Diagonal & Vega=0 & Off-Diagonal & Program ID \\
 &  &  &  &  &  Uncertainty &  Zeropoint & Covariance &  \\
\hline
\hline
04-Mar-12  &  55990.313  &  F475W  &  1032.0  &  0.533  &  0.089  &  26.154  &  \nodata  &  12454 \\
18-Mar-12  &  56004.743  &  F475W  &  1032.0  &  0.122  &  0.077  &  26.154  &  \nodata  &  12454 \\
18-Feb-12  &  55975.698  &  F606W  &  998.0  &  7.927  &  0.197  &  26.407  &  \nodata  &  12454 \\
16-Mar-12  &  56002.615  &  F606W  &  1032.0  &  2.484  &  0.109  &  26.407  &  \nodata  &  12454 \\
03-Feb-12  &  55960.646  &  F625W  &  1032.0  &  0.433  &  0.086  &  25.736  &  \nodata  &  12454 \\
04-Mar-12  &  55990.246  &  F625W  &  1032.0  &  4.232  &  0.127  &  25.736  &  \nodata  &  12454 \\
18-Feb-12  &  55975.682  &  F775W  &  1032.0  &  8.315  &  0.195  &  25.274  &  \nodata  &  12454 \\
04-Mar-12  &  55990.329  &  F775W  &  1013.0  &  8.401  &  0.204  &  25.274  &  \nodata  &  12454 \\
03-Mar-12  &  55989.847  &  F814W  &  1032.0  &  10.237  &  0.228  &  25.523  &  \nodata  &  12454 \\
16-Mar-12  &  56002.631  &  F814W  &  984.0  &  7.234  &  0.184  &  25.523  &  \nodata  &  12454 \\
29-Mar-12  &  56015.541  &  F814W  &  1017.0  &  3.609  &  0.139  &  25.523  &  \nodata  &  12454 \\
12-Apr-12  &  56029.642  &  F814W  &  985.0  &  1.526  &  0.114  &  25.523  &  \nodata  &  12454 \\
03-Feb-12  &  55960.662  &  F850LP  &  1017.0  &  0.370  &  0.073  &  23.900  &  \nodata  &  12454 \\
04-Mar-12  &  55990.262  &  F850LP  &  1017.0  &  2.534  &  0.093  &  23.900  &  \nodata  &  12454 \\
18-Mar-12  &  56004.759  &  F850LP  &  1001.0  &  1.738  &  0.080  &  23.900  &  \nodata  &  12454 \\
12-Apr-12  &  56029.626  &  F850LP  &  1032.0  &  0.670  &  0.063  &  23.900  &  \nodata  &  12454 \\
16-Mar-12  &  56002.683  &  F105W  &  1508.801514  &  9.873  &  0.175  &  25.600  &  \nodata  &  12454 \\
04-Mar-12  &  55990.381  &  F110W  &  1508.801514  &  17.571  &  0.252  &  26.052  &  0.01090  &  12454 \\
12-Apr-12  &  56029.710  &  F110W  &  1005.867676  &  7.492  &  0.302  &  26.052  &  0.01090  &  12454 \\
16-Mar-12  &  56002.700  &  F140W  &  1005.867676  &  5.314  &  0.228  &  25.371  &  \nodata  &  12454 \\
04-Mar-12  &  55990.398  &  F160W  &  1005.867676  &  4.638  &  0.207  &  24.680  &  0.00659  &  12454 \\
12-Apr-12  &  56029.694  &  F160W  &  1508.801514  &  3.658  &  0.168  &  24.680  &  0.00659  &  12454 \\
\hline
\end{tabular}
\caption{Photometry of SN H1.\label{table:h1}}
\end{table*}

\begin{table*}
\begin{tabular}{l  l   l   l  r  r  r  r  r}
\hline
UT Date & MJD & Filter & Exp. Time & Flux & Diagonal & Vega=0 & Off-Diagonal & Program ID \\
 &  &  &  &  &  Uncertainty &  Zeropoint & Covariance & \\
\hline
\hline
18-Jan-11  &  55579.356  &  F606W  &  1032.0  &  0.882  &  0.144  &  26.407  &  \nodata  &  12065 \\
22-Jan-11  &  55583.433  &  F606W  &  1073.0  &  0.595  &  0.131  &  26.407  &  \nodata  &  12065 \\
19-Nov-11  &  55884.956  &  F606W  &  2254.0  &  0.000  &  0.090  &  26.407  &  \nodata  &  12099 \\
18-Nov-10  &  55518.913  &  F625W  &  1032.0  &  0.000  &  0.094  &  25.736  &  \nodata  &  12065 \\
04-Jan-11  &  55565.975  &  F625W  &  1032.0  &  1.538  &  0.104  &  25.736  &  \nodata  &  12065 \\
18-Nov-10  &  55518.995  &  F775W  &  1010.0  &  0.000  &  0.111  &  25.274  &  \nodata  &  12065 \\
22-Jan-11  &  55583.416  &  F775W  &  1032.0  &  1.543  &  0.119  &  25.274  &  \nodata  &  12065 \\
08-Dec-10  &  55538.433  &  F814W  &  1060.0  &  0.000  &  0.128  &  25.523  &  \nodata  &  12065 \\
28-Dec-10  &  55558.470  &  F814W  &  1092.0  &  2.287  &  0.122  &  25.523  &  \nodata  &  12065 \\
18-Jan-11  &  55579.373  &  F814W  &  1059.0  &  3.024  &  0.143  &  25.523  &  \nodata  &  12065 \\
07-Feb-11  &  55599.391  &  F814W  &  1032.0  &  1.518  &  0.128  &  25.523  &  \nodata  &  12065 \\
18-Nov-10  &  55518.929  &  F850LP  &  1014.0  &  -0.002  &  0.071  &  23.900  &  \nodata  &  12065 \\
08-Dec-10  &  55538.417  &  F850LP  &  1032.0  &  0.002  &  0.065  &  23.900  &  \nodata  &  12065 \\
04-Jan-11  &  55565.991  &  F850LP  &  1092.0  &  0.783  &  0.069  &  23.900  &  \nodata  &  12065 \\
21-Feb-11  &  55613.178  &  F850LP  &  1994.0  &  0.275  &  0.044  &  23.900  &  \nodata  &  12099 \\
01-Mar-11  &  55621.441  &  F850LP  &  1076.0  &  -0.006  &  0.069  &  23.900  &  \nodata  &  12065 \\
24-Jan-11  &  55585.083  &  F105W  &  805.9  &  5.414  &  0.284  &  25.600  &  \nodata  &  12360 \\
24-Jan-11  &  55585.116  &  F125W  &  805.9  &  5.658  &  0.295  &  25.322  &  \nodata  &  12360 \\
24-Jan-11  &  55585.150  &  F160W  &  905.9  &  3.225  &  0.284  &  24.680  &  \nodata  &  12360 \\
\hline
\end{tabular}
\caption{Photometry of SN A1.\label{table:a1}}
\end{table*}

\begin{table*}
\begin{tabular}{l  l   l   l  r  r  r  r  r}
UT Date & MJD & Filter & Exp. Time & Flux & Diagonal & Vega=0 & Off-Diagonal & Program ID \\
 &  &  &  &  &  Uncertainty &  Zeropoint & Covariance  & \\
\hline
\hline
22-Apr-12  &  56039.072  &  F814W  &  1032.0  &  0.077  &  0.091  &  25.523  &  \nodata  &  12455 \\
22-May-12  &  56069.685  &  F814W  &  1007.0  &  -0.108  &  0.107  &  25.523  &  \nodata  &  12455 \\
17-Jun-12  &  56095.686  &  F814W  &  975.0  &  1.703  &  0.114  &  25.523  &  \nodata  &  12455 \\
26-Mar-12  &  56012.617  &  F850LP  &  1007.0  &  0.062  &  0.058  &  23.900  &  \nodata  &  12455 \\
25-Apr-12  &  56042.014  &  F850LP  &  1007.0  &  -0.086  &  0.063  &  23.900  &  \nodata  &  12455 \\
05-May-12  &  56052.587  &  F850LP  &  991.0  &  0.011  &  0.060  &  23.900  &  \nodata  &  12455 \\
17-Jun-12  &  56095.670  &  F850LP  &  1032.0  &  0.665  &  0.065  &  23.900  &  \nodata  &  12455 \\
22-Apr-12  &  56039.221  &  F105W  &  1305.9  &  \nodata  &  \nodata  &  \nodata  &  \nodata  &  12455 \\
09-May-12  &  56056.030  &  F105W  &  1408.8  &  \nodata  &  \nodata  &  \nodata  &  \nodata  &  12455 \\
02-Jul-12  &  56110.099  &  F105W  &  1005.9  &  5.699  &  0.220  &  25.600  &  0.01112  &  12360 \\
16-Jul-12  &  56124.111  &  F105W  &  1005.9  &  5.170  &  0.259  &  25.600  &  0.01112  &  12360 \\
23-Jul-12  &  56131.379  &  F105W  &  455.9  &  3.539  &  0.282  &  25.600  &  0.01112  &  12360 \\
25-Apr-12  &  56042.132  &  F110W  &  1408.8  &  \nodata  &  \nodata  &  \nodata  &  \nodata  &  12455 \\
17-Jun-12  &  56095.754  &  F110W  &  1005.9  &  7.363  &  0.307  &  26.052  &  \nodata  &  12455 \\
22-Apr-12  &  56039.204  &  F140W  &  1305.9  &  \nodata  &  \nodata  &  \nodata  &  \nodata  &  12455 \\
09-May-12  &  56056.046  &  F140W  &  1005.9  &  \nodata  &  \nodata  &  \nodata  &  \nodata  &  12455 \\
02-Jul-12  &  56110.161  &  F140W  &  1005.9  &  6.877  &  0.275  &  25.371  &  0.01654  &  12360 \\
16-Jul-12  &  56124.175  &  F140W  &  1005.9  &  6.480  &  0.271  &  25.371  &  0.01654  &  12360 \\
26-Mar-12  &  56012.685  &  F160W  &  1005.9  &  \nodata  &  \nodata  &  \nodata  &  \nodata  &  12455 \\
25-Apr-12  &  56042.148  &  F160W  &  1005.9  &  \nodata  &  \nodata  &  \nodata  &  \nodata  &  12455 \\
05-May-12  &  56052.645  &  F160W  &  1408.8  &  \nodata  &  \nodata  &  \nodata  &  \nodata  &  12455 \\
17-Jun-12  &  56095.738  &  F160W  &  1408.8  &  3.275  &  0.193  &  24.680  &  0.00941  &  12455 \\
23-Jul-12  &  56131.444  &  F160W  &  455.9  &  3.218  &  0.312  &  24.680  &  0.00941  &  12360 \\
\hline
\end{tabular}
\caption{Photometry of SN L2.\label{table:l2}}
\end{table*}

\label{lastpage}
\end{document}